%% file: sample-sigconf.tex
\documentclass[sigconf]{acmart}

\usepackage{booktabs} 
\usepackage{todonotes}
\usepackage{amssymb}
\usepackage{enumitem}
\usepackage{tikz}

\begin{document}

\copyrightyear{2018}
\acmYear{2018}
\setcopyright{acmcopyright}
\acmConference[SIGIR '18]{The 41st International ACM SIGIR Conference on Research and Development in Information Retrieval}{July 8--12, 2018}{Ann Arbor, MI, USA}
\acmBooktitle{SIGIR '18: The 41st International ACM SIGIR Conference on Research and Development in Information Retrieval, July 8--12, 2018, Ann Arbor, MI, USA}
\acmPrice{15.00}
\acmDOI{10.1145/3209978.3210135}
\acmISBN{978-1-4503-5657-2/18/07}

\fancyhead{}

\title{Characterizing Question Facets for Complex Answer Retrieval}


\author{Sean MacAvaney}
\affiliation{\institution{IRLab, Georgetown University}}
\email{sean@ir.cs.georgetown.edu}

\author{Andrew Yates}
\affiliation{\institution{Max Planck Institute for Informatics}}
\email{ayates@mpi-inf.mpg.de}

\author{Arman Cohan, Luca Soldaini}
\affiliation{\institution{IRLab, Georgetown University}}
\email{{arman,luca}@ir.cs.georgetown.edu}

\author{Kai Hui}
\affiliation{\institution{SAP SE}}
\email{kai.hui@sap.com}

\author{Nazli Goharian, Ophir Frieder}
\affiliation{\institution{IRLab, Georgetown University}}
\email{{nazli,ophir}@ir.cs.georgetown.edu}

\renewcommand{\shortauthors}{S. MacAvaney et al.}

\begin{abstract}
Complex answer retrieval (CAR) is the process of retrieving answers to questions that have multifaceted or nuanced answers. In this work, we present two novel approaches for CAR based on the observation that question facets can vary in utility: from structural (facets that can apply to many similar topics, such as `History') to topical (facets that are specific to the question's topic, such as the `Westward expansion' of the United States). We first explore a way to incorporate facet utility into ranking models during query term score combination. We then explore a general approach to reform the structure of ranking models to aid in learning of facet utility in the query-document term matching phase. When we use our techniques with a leading neural ranker on the TREC CAR dataset, our methods rank first in the 2017 TREC CAR benchmark, and yield up to 26\% higher performance than the next best method.
\end{abstract}

\maketitle

%
%

\input{intro}

\input{background}

\input{method}
\input{eval}
\input{concl}

\bibliographystyle{ACM-Reference-Format}
\bibliography{sample-bibliography} 

\end{document}

%% file: intro.tex
\section{Introduction} 
\label{sec.introduction}

As people become more comfortable using question answering systems, it is inevitable that they will begin to expect the systems to answer questions with complex answers. For instance, even the seemingly simple question \textit{``Is cheese healthy?''} cannot be answered with a simple `yes' or `no'. To fully answer the question, positive and negative qualities should be discussed, along with the strength of evidence, and conditions under which the qualities apply---a complex answer. Complex Answer Retrieval (CAR) frames this problem as an information retrieval (IR) task~\cite{dietz2017car}. Given a query that consists of a topic (e.g., `cheese'), and facets of the topic (e.g., `health effects'), a CAR system should be able to retrieve information from a variety of sources to throughly answer the corresponding question.

CAR has similarities with existing, yet distinct, areas of research in IR. Although CAR involves passage retrieval, it is distinguishable from passage retrieval because CAR compiles multiple passages together to form complete answers. It is also different than factoid question answering (questions with a simple answer, e.g. \textit{``Who wrote Hamlet?''}), and complex question answering (questions that themselves require reasoning, e.g. \textit{``Which female characters are in the same room as Homer in Act III Scene I?''}).

We observe that question facets can be structural or topical. Structural facets refer to general categories of information that could apply to other entities of the same type, such as the `History' or `Economy' of a country. Topical facets refer to categories of information that are specific to the entity mentioned in the question, such as the `Westward expansion' or `Banking crisis' of the United States. (Although either facet could be asked about other topics, they are much more specific to details of the topic than structural headings.) We call this distinction \textit{facet utility}, and explain it in detail in Section~\ref{sec.background}, along with additional background and related work. We then present and evaluate two novel approaches to CAR based on this observation and the hypothesis that it will affect how terms are matched. The first approach integrates predictors of a facet's utility into the score combination component of an answer ranker. The second approach is a technique to help any model learn to make the distinction itself by treating different facets independently. To predict facet utility, we use the heading structure of CAR queries (described in Section~\ref{sec.background}) and corpus statistics. We show how our approaches can be integrated with recent neural ranking models, and evaluate on the TREC CAR dataset. Our approaches yield favorable results compared to other known methods, achieving the top results overall and up to a 26\% gain over the next best method.

%% file: background.tex
\section{Background and Related Work}\label{sec.background}

The first major work done with CAR frames the task in terms of Wikipedia content generation~\cite{dietz2017cardata}. CAR fits naturally with this domain because CAR query topics and facets often correspond well with article titles and headings, respectively. Furthermore, Wikipedia itself provides an extensive source of sample queries (paths in the heading hierarchy from the title), partial answers (i.e., paragraphs), and automatic relevance judgments (paragraphs can be assumed relevant to the headings they are under). For simplicity, we use Wikipedia-focused terminology in the remainder of this work. A \textit{heading} refers to any component of a query, and corresponds to a question topic or facet. The \textit{title} is the first query component (topic), the \textit{main heading} is the last component, and \textit{intermediate heading} are any headings between the two (if any). The main and intermediate headings represent the facet of interest to the topic. Example queries using this terminology are given in Table~\ref{tab:ex-queries}.

A central challenge of CAR is resolving \textit{facet utility}. Due to the structure of CAR queries as a list of headings, we generalize the concept to \textit{heading utility}---the idea that headings (i.e., question topics and facets) can serve a variety of functions in an article. We distinguish between structural and topical headings. We define \textit{structural headings} as headings that serve a structural purpose for an article---general question facets that could be asked about many similar topics. In contrast, \textit{topical headings} describe details that are specific to the particular topic. For instance, \textit{``cooking and eating''} is a structural heading for Cheese (one would expect it to be found in other food-related articles), whereas \textit{``cheeseboard''} is a topical heading because it relates specifically to the topic of the article. Because the terminology in structural headings is necessarily more generic (they accommodate many topics), we predict that terms found in these headings are less likely to appear verbatim in relevant paragraphs than terms in topical headings. Thus, modeling this behavior should improve performance on CAR because it will be able to learn which terms are less important. Previous work does not model facet utility, treating all headings equally by concatenating their components.

\begin{table}
\scriptsize
\setlength{\tabcolsep}{0.5em}
\caption{Example CAR queries from Wikipedia by heading position. Some queries have no intermediate headings.}\label{tab:ex-queries}
\begin{tabular}{lllll}
\toprule
Title & & Intermediate Heading(s) & & Main Heading \\
\midrule
Cheese &\guillemotright{}& \textit{(none)} &\guillemotright{}& Nutrition and health\\
Green sea turtle & \guillemotright & Ecology and behavior & \guillemotright & Life cycle \\
History of the United States & \guillemotright & 20th Century & \guillemotright & Imperialism \\
Disturbance (ecology) & \guillemotright & \textit{(none)} & \guillemotright & Cyclic disturbance \\
Medical tourism & \guillemotright & Destinations \guillemotright{} Europe & \guillemotright & Finland \\
\bottomrule
\end{tabular}
\vspace{-2em}
\end{table}


\citet{DBLP:journals/corr/NanniMMD17} presents a survey of prominent general domain ranking and query expansion approaches for CAR. They test one deep neural model (Duet~\cite{mitra2017learning}), and find that it outperforms the other approaches, including BM25, cosine similarity with TF-IDF and word embeddings, and a learning-to-rank approach. The recent 2017 TREC track focused on CAR~\cite{dietz2017car}. This track yielded both manual relevance judgments for evaluation of CAR systems, and a variety of new CAR approaches (seven teams participated). One prominent approach used a sequential dependence model~\cite{cuis2017}. They modified the approach for CAR by limiting ordered ngrams to those found within a single heading, and unordered ngrams to only inter-heading pairs. Another approach uses a Siamese attention network~\cite{utd2017}, including topic features extracted from DBPedia. While this approach does distinguish the title from other headings, it only uses it for query expansion and related entity extraction. Another submission applied a reinforcement learning-based query reformulation approach to CAR~\cite{nyudl2017}.

%% file: method.tex
\begin{figure*}
\includegraphics[scale=0.9]{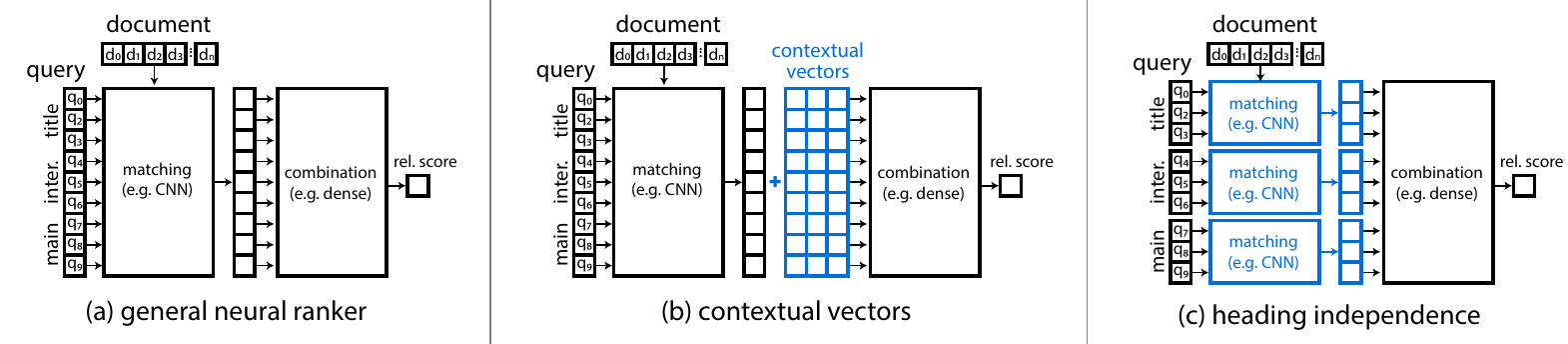}
\caption{(a) General interaction-focused ranking architecture, with matching and combination phases (unmodified). (b) Modified architecture, including contextual vectors for combination. (c) Modified architecture, splitting for heading independence.}\label{fig.model}
\end{figure*}

\section{Method}\label{sec.method}

Since previous work shows that neural-based rankers have potential for CAR, we focus on an approach that can be adapted for various neural rankers. Many leading interaction-focused neural rankers share a similar two-phase architecture, as shown in Figure~\ref{fig.model}a. Phase 1 performs matching of query terms to document terms, and phase 2 combines the matching results to produce a final relevance score. For instance, DRMM~\cite{guo2016deep} uses a feed-forward histogram matching network, and a term gating combination network to predict relevance. MatchPyramid~\cite{DBLP:journals/corr/PangLGXC16} uses hierarchal convolution for matching, followed by a dense layer for aggregation. Similarly, PACRR~\cite{hui2017position} uses a max-pooled convolution phase for matching, and a recurrent or dense combination phase. Finally, DeepRank~\cite{pang2017deeprank} generates query contexts and uses a convolutional layer to generate local relevance representations as a matching phase, and uses a term gating mechanism for combination.
We present two approaches to model facet utility by modifying this generalized neural ranking structure. The first approach applies \textit{contextual vectors} in the combination phase (Figure~\ref{fig.model}b), and the second approach splits the input into independent matching phases (Figure~\ref{fig.model}c).

\textbf{Contextual vectors.}
In the combination phase, signals across query terms are combined to produce a relevance score, so it is natural to include information here to provide additional context about each query term when combining the results. For instance, PACRR includes the inverse document frequency (IDF) in its combination layer, allowing the model to learn how to weight results based on this statistic~\cite{hui2017position}. We use this phase to inform the model about heading utility based on predictions about the distinction between structural and topical headings. We call these \textit{contextual vectors}, since they provide context in the CAR domain. The intuition is that by providing the model with estimators of heading utility, the model will learn which terms to weight higher. Here we explore two types of contextual vectors: \textit{heading position}~(HP) and \textit{heading frequency}~(HF).

When distinguishing between structural and topical headings, it is important to consider the position itself in the query. For instance, since the title is the question topic, it is necessarily topical. Furthermore, it is reasonable to suspect that intermediate headings will often be structural because they assist in the organization of an article. Main headings may either be structural or topical, depending on the question itself. Thus, for \textit{heading position} contextual vectors, we use a simple indicator to distinguish whether a term is from the title, an intermediate, or the main heading. An example is given in Figure~\ref{tab:vecs}.

\begin{table}
\scriptsize
\setlength\tabcolsep{0.5em}
\caption{Example contextual vectors for the query ``\textit{green sea turtle \guillemotright{} ecology and behavior \guillemotright{} life cycle}''.}\label{tab:vecs}
\centering
\begin{tabular}{l|ccc|ccc|cc}
\toprule
 & green & sea & turtle & ecology & and & behavior & life & cycle \\
\midrule
position\_title    & 1 & 1 & 1 & 0 & 0 & 0 & 0 & 0 \\
position\_inter    & 0 & 0 & 0 & 1 & 1 & 1 & 0 & 0 \\
position\_main     & 0 & 0 & 0 & 0 & 0 & 0 & 1 & 1  \\
\midrule
heading\_frequency & 0 & 0 & 0 & 3 & 3 & 3 & 3 & 3 \\
\bottomrule
\end{tabular}
\vspace{-1em}
\end{table}

Another approach to modeling structural and topical headings using contextual vectors is to examine the prevalence of a given heading. This is based on the intuition that structural headings should appear in many similar documents, whereas the usage of topical headings should be less widespread. For instance, the structural heading \textit{``Nutrition and health''} in the article \textit{Cheese} also appears in articles entitled \textit{Beef}, \textit{Raisin}, \textit{Miso}, and others, whereas the topical \textit{``Cheeseboard''} heading only also appears as the title of a disambiguation page. We model this behavior using \textit{heading usage frequency}:
$
frq(h)=\frac
  {\sum_{a\in C}{I(h \in a)}}
  {|C|}
$.
That is, the probability that a given article $a$ in corpus $C$ contains heading $h$, given the indicator function $I$. Heading usage frequencies very close to 0 include titles and other content-specific headings like \textit{Cheeseboard}. Due to the wide variety of Wikipedia articles, most probabilities are very low. Therefore, we stratify the scores by percentile, grouping similarly-common headings together. Based on pilot studies, we found the (1) 60th, (2) 90th, and (3) 99th percentiles to be effective breakpoints.
We use complete, case insensitive heading matches.
Unknown headings are assumed to be infrequent, and belong to the 0th percentile.
An example of this vector is given in Table~\ref{tab:vecs}.

\textbf{Heading independence.}
Since contextual vectors are applied in the combination phase, they have no effect on the criteria constituting a strong signal from the matching phase. However, we hypothesize that facet utility can also be important when matching. For instance, a structural heading like \textit{``History''} might have a lower matching threshold, allowing matches of similar words terms such as \textit{``early''} or \textit{``was''} (both of which have a lower \textsc{word2vec} cosine similarity score to \textit{``history''} than functionally-related word pairs, such as \textit{``cheese''} and \textit{``chocolate''}).

Thus, we propose a method called~\textit{heading independence}. With this approach, we modify the structure of a generic neural IR model by splitting the matching stage into three independent parts: one for the title, one for intermediate headings, and one for the main heading. Each sub-matching phase operates independently as it otherwise would for the combined query. Then, the results are combined using the same combination logic of the original model (e.g., a dense or recurrent layer). This allows the model to learn separated logic for different heading components. The reasoning behind the split by query component is the same as the reasoning behind using heading position vectors: the title is topical, whereas intermediate headings are likely structural, and the main heading could be either. With separate matching logic for each, the model should be able to more easily distinguish between the types.

An added benefit of this approach is that it improves heading alignment in the combination phase. When headings are simply concatenated (even with a symbol to indicate a change in headings), the alignment of each query component will vary among queries. Since the output of each matching stage is fixed in size, the locations of each query component will be consistent among queries. We suspect that this is particularly useful when using dense combination. 


%% file: eval.tex

\begin{table*}
\small
\caption{Performance results on \textit{benchmarkY1test}. The top value is in bold. Records marked with * are based on official TREC runs, and had top results included in the manual assessment pool. Significant results compared to the unmodified PACRR model are marked with $\blacktriangle$ and $\blacktriangledown$ (paired t-test, 95\% confidence). The abbreviations for our methods are as follows: HP is the heading position contextual vector; HF is the heading frequency contextual vector; HI is heading independence.}\label{tab:results}
\begin{tabular}{lrrrrrrrr}
\toprule
& \multicolumn{4}{c}{Automatic} & \multicolumn{4}{c}{Manual} \\
\cmidrule(lr){2-5}\cmidrule(lr){6-9}
Approach & MAP & R-Prec & MRR & nDCG & MAP & R-Prec & MRR & nDCG \\
\midrule
PACRR (no modification) &  0.164 &  0.131 &  0.247 &  0.254 &  0.208 &  0.219 &  0.445 &  0.403 \\
PACRR + HP* & $\blacktriangle$ 0.170 &  0.135 & $\blacktriangle$ 0.258 & $\blacktriangle$ 0.260 &  0.209 &  0.218 &  0.452 &  0.406 \\
PACRR + HP + HF* & $\blacktriangle$ 0.170 &  0.134 & $\blacktriangle$ 0.255 & $\blacktriangle$ 0.259 & \bf$\blacktriangle$ 0.211 & \bf 0.221 & \bf 0.453 & \bf$\blacktriangle$ 0.408 \\
PACRR + HI & $\blacktriangle$ 0.171 &  0.139 & $\blacktriangle$ 0.256 & $\blacktriangle$ 0.260 &  0.205 &  0.213 &  0.442 &  0.403 \\
PACRR + HI + HF & \bf$\blacktriangle$ 0.176 & \bf$\blacktriangle$ 0.146 & \bf$\blacktriangle$ 0.263 & \bf$\blacktriangle$ 0.265 &  0.204 &  0.214 &  0.440 &  0.401 \\
\midrule
Sequential dependence model*~\cite{cuis2017} & $\blacktriangledown$ 0.150 & $\blacktriangledown$ 0.116 & $\blacktriangledown$ 0.226 & $\blacktriangledown$ 0.238 & $\blacktriangledown$ 0.172 & $\blacktriangledown$ 0.186 & $\blacktriangledown$ 0.393 & $\blacktriangledown$ 0.350 \\
Siamese attention network*~\cite{utd2017} & $\blacktriangledown$ 0.121 & $\blacktriangledown$ 0.096 & $\blacktriangledown$ 0.185 & $\blacktriangledown$ 0.175 & $\blacktriangledown$ 0.137 & $\blacktriangledown$ 0.171 & $\blacktriangledown$ 0.345 & $\blacktriangledown$ 0.274 \\
BM25 baseline* & $\blacktriangledown$ 0.122 & $\blacktriangledown$ 0.097 & $\blacktriangledown$ 0.183 & $\blacktriangledown$ 0.196 & $\blacktriangledown$ 0.138 & $\blacktriangledown$ 0.158 & $\blacktriangledown$ 0.317 & $\blacktriangledown$ 0.296 \\
\bottomrule
\end{tabular}
\end{table*}


\section{Experimental setup}
\textbf{Dataset.}
TREC CAR provides several sets of queries based on a recent dump of Wikipedia~\cite{dietz2017cardata,dietz2017car}.
Queries in each set are generated from the heading structure of an article, where each query represents a path from the article title down to the main heading. Each query also includes automatic relevance judgments based on the assumption that paragraphs under a given heading are relevant to the query with that main heading.
Half of the dump belongs to the \texttt{train} set, which is split into 5 folds. We use folds 1 and 2 in this work, consisting of $873,746$ queries and $2.2M$ automatic relevance judgments (more data than this was not required for our models to converge).
The \texttt{test200} set contains 1,860 queries and $4.7k$ automatic relevance judgments.
The \texttt{benchmarkY1test} set contains 2,125 queries and $5.8k$ automatic relevance judgments. It also includes $30k$ manual relevance judgments, ranging from \textit{Trash} (-2) to \textit{Must be mentioned} (3).
The \texttt{paragraphcorpus} is a collection of $30M$ paragraphs from the Wikipedia dump with no article or heading structure provided, functioning as a source of answers for retrieval.

\textbf{Model integration.}
We evaluate our contextual vector and heading independence approaches using the Position-Aware Convolutional Recurrent Relevance neural IR architecture (PACRR)~\cite{hui2017position}, which is a strong neural retrieval model with a structure that naturally lends itself to incorporating contextual vectors and heading independence signals. We refer the reader to~\citet{hui2017position} for full details about the model, but we give a short description here to provide details about how our approach is integrated. PACRR first processes square convolutional filters over a $q\times d$ query-document similarity matrix, where each cell represents similarity scores between the corresponding query and document term. The filters are max-pooled for each cell, and the scores are k-max pooled over each query term ($k=2$). Then a dense layer combines the scores (along with term IDF scores) to yield a final relevance score for the query-document pair. For runs that include \textit{contextual vectors}, we append them to each term (alongside IDF) during combination. For \textit{heading independence}, we use separate convolution and pooling layers, followed by a dense layer for each heading component. We also explore using the heading frequency contextual vector when using heading independence (included after the pooling layer), and before the independent dense layer.

\textbf{Training and evaluation.}
We train the models on samples from \texttt{train.fold1} and \texttt{train.fold2}. Positive training examples come from the automatic relevance judgments, whereas negative training examples are selected from the top non-relevant BM25 results for the given query. Each model is trained for 80 iterations, and the top training iteration is selected using the R-Prec on \texttt{test200}. Evaluation is conducted with automatic and manual judgments on \texttt{benchmarkY1test}. The results are based on an initial ranking of the top 100 BM25 results for each query. We report Mean Average Precision (MAP), R-Precision (R-Prec), Mean Reciprocal Rank (MRR), and normalized Discounted Cumulative Gain (nDCG) of each variation (all four official TREC CAR metrics).

\section{Results}
We present system performance in Table~\ref{tab:results}. Our methods are compared to the unmodified PACRR model, two other top submissions to TREC CAR 2017 (sequential dependency model~\cite{cuis2017} and the Siamese attention network~\cite{utd2017}), and a BM25 baseline (which produces the initial result set that our methods re-rank).

Our method outperforms the other TREC submissions and the BM25 baseline by all metrics for both manual and automatic relevance judgments (paired t-test, 95\% confidence). The method that uses heading independence (HI) and the heading frequency vector (HF) yields up to a 26\% improvement over the next best approach (SDM).

Our approach also consistently outperforms the unmodified version of PACRR when evaluating using automatic relevance judgments, performing up to 11\% better than the unmodified version of PACRR. Our approach occasionally does better than unmodified PACRR when evaluating with manual relevance judgments. Specifically, our approach that uses the heading position (HP) and heading frequency (HF) contextual vectors does the best overall. We acknowledge that this method (and the version with only heading position) were included as official TREC runs, yielding an advantage in the manual comparison.

\begin{figure}
\scalebox{0.6}{\input{comp}}
{\par\vspace{-0.5em}\small\centering Term occurrence rate}
\caption{Kernel density estimation for main (solid), intermediate (dashed), and title (dotted) heading term occurrence rates, based on automatic judgments in train.fold0.}\label{fig.comp}
\vspace{-1em}
\end{figure}
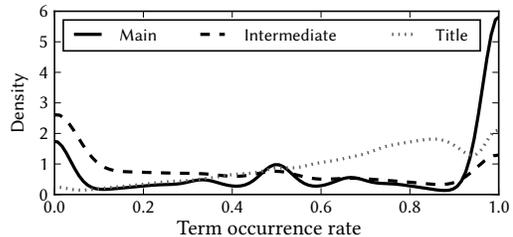

This work is based on the distinction between structural and topical headings, and the differences in how they interact with terms in relevant documents. While there is no absolute distinction between the two, we presented various approaches to approximate the distinction. By plotting the \textit{term occurrence rate} (that is, the probability that any term occurs in a relevant paragraph) for title, intermediate, and main headings, we see clear differences in the distribution (Figure~\ref{fig.comp}). Particularly, the plot shows that main headings are much more likely to appear in relevant documents than title and intermediate headings. Furthermore, the distributions of intermediate and title headings are roughly opposite each other, with titles (topical) more likely to occur than intermediate headings (structural).

%% file: comp.tex
\begingroup%
\makeatletter%
\begin{pgfpicture}%
\pgfpathrectangle{\pgfpointorigin}{\pgfqpoint{5.000000in}{2.000000in}}%
\pgfusepath{use as bounding box, clip}%
\begin{pgfscope}%
\pgfsetbuttcap%
\pgfsetmiterjoin%
\definecolor{currentfill}{rgb}{1.000000,1.000000,1.000000}%
\pgfsetfillcolor{currentfill}%
\pgfsetlinewidth{0.000000pt}%
\definecolor{currentstroke}{rgb}{1.000000,1.000000,1.000000}%
\pgfsetstrokecolor{currentstroke}%
\pgfsetdash{}{0pt}%
\pgfpathmoveto{\pgfqpoint{0.000000in}{0.000000in}}%
\pgfpathlineto{\pgfqpoint{5.000000in}{0.000000in}}%
\pgfpathlineto{\pgfqpoint{5.000000in}{2.000000in}}%
\pgfpathlineto{\pgfqpoint{0.000000in}{2.000000in}}%
\pgfpathclose%
\pgfusepath{fill}%
\end{pgfscope}%
\begin{pgfscope}%
\pgfsetbuttcap%
\pgfsetmiterjoin%
\definecolor{currentfill}{rgb}{1.000000,1.000000,1.000000}%
\pgfsetfillcolor{currentfill}%
\pgfsetlinewidth{0.000000pt}%
\definecolor{currentstroke}{rgb}{0.000000,0.000000,0.000000}%
\pgfsetstrokecolor{currentstroke}%
\pgfsetstrokeopacity{0.000000}%
\pgfsetdash{}{0pt}%
\pgfpathmoveto{\pgfqpoint{0.625000in}{0.200000in}}%
\pgfpathlineto{\pgfqpoint{4.500000in}{0.200000in}}%
\pgfpathlineto{\pgfqpoint{4.500000in}{1.800000in}}%
\pgfpathlineto{\pgfqpoint{0.625000in}{1.800000in}}%
\pgfpathclose%
\pgfusepath{fill}%
\end{pgfscope}%
\begin{pgfscope}%
\pgfpathrectangle{\pgfqpoint{0.625000in}{0.200000in}}{\pgfqpoint{3.875000in}{1.600000in}} %
\pgfusepath{clip}%
\pgfsetrectcap%
\pgfsetroundjoin%
\pgfsetlinewidth{2.007500pt}%
\definecolor{currentstroke}{rgb}{0.000000,0.000000,0.000000}%
\pgfsetstrokecolor{currentstroke}%
\pgfsetdash{}{0pt}%
\pgfpathmoveto{\pgfqpoint{0.615000in}{0.664374in}}%
\pgfpathlineto{\pgfqpoint{0.616556in}{0.665394in}}%
\pgfpathlineto{\pgfqpoint{0.653622in}{0.658372in}}%
\pgfpathlineto{\pgfqpoint{0.690688in}{0.621829in}}%
\pgfpathlineto{\pgfqpoint{0.727753in}{0.563338in}}%
\pgfpathlineto{\pgfqpoint{0.801884in}{0.424338in}}%
\pgfpathlineto{\pgfqpoint{0.838950in}{0.363453in}}%
\pgfpathlineto{\pgfqpoint{0.876016in}{0.315851in}}%
\pgfpathlineto{\pgfqpoint{0.913081in}{0.282483in}}%
\pgfpathlineto{\pgfqpoint{0.950147in}{0.261657in}}%
\pgfpathlineto{\pgfqpoint{0.987212in}{0.250431in}}%
\pgfpathlineto{\pgfqpoint{1.024278in}{0.245751in}}%
\pgfpathlineto{\pgfqpoint{1.061343in}{0.245105in}}%
\pgfpathlineto{\pgfqpoint{1.098409in}{0.246726in}}%
\pgfpathlineto{\pgfqpoint{1.172540in}{0.252866in}}%
\pgfpathlineto{\pgfqpoint{1.469065in}{0.280636in}}%
\pgfpathlineto{\pgfqpoint{1.543196in}{0.285136in}}%
\pgfpathlineto{\pgfqpoint{1.654393in}{0.288832in}}%
\pgfpathlineto{\pgfqpoint{1.691459in}{0.291586in}}%
\pgfpathlineto{\pgfqpoint{1.728524in}{0.296416in}}%
\pgfpathlineto{\pgfqpoint{1.765590in}{0.303446in}}%
\pgfpathlineto{\pgfqpoint{1.839721in}{0.320230in}}%
\pgfpathlineto{\pgfqpoint{1.876787in}{0.326410in}}%
\pgfpathlineto{\pgfqpoint{1.913852in}{0.328755in}}%
\pgfpathlineto{\pgfqpoint{1.950918in}{0.326448in}}%
\pgfpathlineto{\pgfqpoint{1.987983in}{0.319831in}}%
\pgfpathlineto{\pgfqpoint{2.025049in}{0.310224in}}%
\pgfpathlineto{\pgfqpoint{2.099180in}{0.289148in}}%
\pgfpathlineto{\pgfqpoint{2.136246in}{0.280692in}}%
\pgfpathlineto{\pgfqpoint{2.173311in}{0.274955in}}%
\pgfpathlineto{\pgfqpoint{2.210377in}{0.272720in}}%
\pgfpathlineto{\pgfqpoint{2.247442in}{0.274978in}}%
\pgfpathlineto{\pgfqpoint{2.284508in}{0.283032in}}%
\pgfpathlineto{\pgfqpoint{2.321574in}{0.298198in}}%
\pgfpathlineto{\pgfqpoint{2.358639in}{0.321139in}}%
\pgfpathlineto{\pgfqpoint{2.395705in}{0.351048in}}%
\pgfpathlineto{\pgfqpoint{2.469836in}{0.418443in}}%
\pgfpathlineto{\pgfqpoint{2.506902in}{0.445271in}}%
\pgfpathlineto{\pgfqpoint{2.543967in}{0.460210in}}%
\pgfpathlineto{\pgfqpoint{2.581033in}{0.460050in}}%
\pgfpathlineto{\pgfqpoint{2.618098in}{0.444792in}}%
\pgfpathlineto{\pgfqpoint{2.655164in}{0.417653in}}%
\pgfpathlineto{\pgfqpoint{2.729295in}{0.349735in}}%
\pgfpathlineto{\pgfqpoint{2.766361in}{0.319685in}}%
\pgfpathlineto{\pgfqpoint{2.803426in}{0.296754in}}%
\pgfpathlineto{\pgfqpoint{2.840492in}{0.281825in}}%
\pgfpathlineto{\pgfqpoint{2.877558in}{0.274337in}}%
\pgfpathlineto{\pgfqpoint{2.914623in}{0.273100in}}%
\pgfpathlineto{\pgfqpoint{2.951689in}{0.276946in}}%
\pgfpathlineto{\pgfqpoint{2.988754in}{0.284979in}}%
\pgfpathlineto{\pgfqpoint{3.025820in}{0.296404in}}%
\pgfpathlineto{\pgfqpoint{3.137017in}{0.337168in}}%
\pgfpathlineto{\pgfqpoint{3.174082in}{0.345872in}}%
\pgfpathlineto{\pgfqpoint{3.211148in}{0.348887in}}%
\pgfpathlineto{\pgfqpoint{3.248213in}{0.345816in}}%
\pgfpathlineto{\pgfqpoint{3.285279in}{0.337778in}}%
\pgfpathlineto{\pgfqpoint{3.359410in}{0.316174in}}%
\pgfpathlineto{\pgfqpoint{3.396476in}{0.307259in}}%
\pgfpathlineto{\pgfqpoint{3.433541in}{0.301222in}}%
\pgfpathlineto{\pgfqpoint{3.470607in}{0.297814in}}%
\pgfpathlineto{\pgfqpoint{3.581804in}{0.292356in}}%
\pgfpathlineto{\pgfqpoint{3.655935in}{0.284911in}}%
\pgfpathlineto{\pgfqpoint{3.767132in}{0.268940in}}%
\pgfpathlineto{\pgfqpoint{3.952460in}{0.241242in}}%
\pgfpathlineto{\pgfqpoint{3.989525in}{0.237116in}}%
\pgfpathlineto{\pgfqpoint{4.026591in}{0.235073in}}%
\pgfpathlineto{\pgfqpoint{4.063657in}{0.237232in}}%
\pgfpathlineto{\pgfqpoint{4.100722in}{0.247345in}}%
\pgfpathlineto{\pgfqpoint{4.137788in}{0.271415in}}%
\pgfpathlineto{\pgfqpoint{4.174853in}{0.317938in}}%
\pgfpathlineto{\pgfqpoint{4.211919in}{0.397199in}}%
\pgfpathlineto{\pgfqpoint{4.248984in}{0.519052in}}%
\pgfpathlineto{\pgfqpoint{4.286050in}{0.689076in}}%
\pgfpathlineto{\pgfqpoint{4.323116in}{0.903885in}}%
\pgfpathlineto{\pgfqpoint{4.397247in}{1.390607in}}%
\pgfpathlineto{\pgfqpoint{4.434312in}{1.595804in}}%
\pgfpathlineto{\pgfqpoint{4.471378in}{1.725981in}}%
\pgfpathlineto{\pgfqpoint{4.508444in}{1.755442in}}%
\pgfpathlineto{\pgfqpoint{4.510000in}{1.752193in}}%
\pgfpathlineto{\pgfqpoint{4.510000in}{1.752193in}}%
\pgfusepath{stroke}%
\end{pgfscope}%
\begin{pgfscope}%
\pgfpathrectangle{\pgfqpoint{0.625000in}{0.200000in}}{\pgfqpoint{3.875000in}{1.600000in}} %
\pgfusepath{clip}%
\pgfsetbuttcap%
\pgfsetroundjoin%
\pgfsetlinewidth{2.007500pt}%
\definecolor{currentstroke}{rgb}{0.000000,0.000000,0.000000}%
\pgfsetstrokecolor{currentstroke}%
\pgfsetdash{{6.000000pt}{6.000000pt}}{0.000000pt}%
\pgfpathmoveto{\pgfqpoint{0.615000in}{0.887879in}}%
\pgfpathlineto{\pgfqpoint{0.630185in}{0.895892in}}%
\pgfpathlineto{\pgfqpoint{0.670026in}{0.892675in}}%
\pgfpathlineto{\pgfqpoint{0.709868in}{0.866738in}}%
\pgfpathlineto{\pgfqpoint{0.749709in}{0.822169in}}%
\pgfpathlineto{\pgfqpoint{0.789551in}{0.764812in}}%
\pgfpathlineto{\pgfqpoint{0.869234in}{0.637524in}}%
\pgfpathlineto{\pgfqpoint{0.909076in}{0.578731in}}%
\pgfpathlineto{\pgfqpoint{0.948917in}{0.528128in}}%
\pgfpathlineto{\pgfqpoint{0.988759in}{0.487273in}}%
\pgfpathlineto{\pgfqpoint{1.028600in}{0.456226in}}%
\pgfpathlineto{\pgfqpoint{1.068442in}{0.433979in}}%
\pgfpathlineto{\pgfqpoint{1.108283in}{0.418931in}}%
\pgfpathlineto{\pgfqpoint{1.148125in}{0.409312in}}%
\pgfpathlineto{\pgfqpoint{1.187966in}{0.403473in}}%
\pgfpathlineto{\pgfqpoint{1.227808in}{0.400056in}}%
\pgfpathlineto{\pgfqpoint{1.307491in}{0.396753in}}%
\pgfpathlineto{\pgfqpoint{1.586382in}{0.388549in}}%
\pgfpathlineto{\pgfqpoint{1.705907in}{0.386093in}}%
\pgfpathlineto{\pgfqpoint{1.865273in}{0.384022in}}%
\pgfpathlineto{\pgfqpoint{1.944956in}{0.379703in}}%
\pgfpathlineto{\pgfqpoint{2.024639in}{0.372154in}}%
\pgfpathlineto{\pgfqpoint{2.104322in}{0.363758in}}%
\pgfpathlineto{\pgfqpoint{2.144164in}{0.360592in}}%
\pgfpathlineto{\pgfqpoint{2.184005in}{0.358902in}}%
\pgfpathlineto{\pgfqpoint{2.223847in}{0.359166in}}%
\pgfpathlineto{\pgfqpoint{2.263688in}{0.361659in}}%
\pgfpathlineto{\pgfqpoint{2.303530in}{0.366371in}}%
\pgfpathlineto{\pgfqpoint{2.343371in}{0.372959in}}%
\pgfpathlineto{\pgfqpoint{2.462896in}{0.395828in}}%
\pgfpathlineto{\pgfqpoint{2.502738in}{0.400918in}}%
\pgfpathlineto{\pgfqpoint{2.542579in}{0.403118in}}%
\pgfpathlineto{\pgfqpoint{2.582421in}{0.401934in}}%
\pgfpathlineto{\pgfqpoint{2.622262in}{0.397352in}}%
\pgfpathlineto{\pgfqpoint{2.662104in}{0.389848in}}%
\pgfpathlineto{\pgfqpoint{2.741787in}{0.369769in}}%
\pgfpathlineto{\pgfqpoint{2.821470in}{0.350206in}}%
\pgfpathlineto{\pgfqpoint{2.861312in}{0.342856in}}%
\pgfpathlineto{\pgfqpoint{2.901153in}{0.337740in}}%
\pgfpathlineto{\pgfqpoint{2.940995in}{0.334902in}}%
\pgfpathlineto{\pgfqpoint{2.980836in}{0.334093in}}%
\pgfpathlineto{\pgfqpoint{3.060519in}{0.336595in}}%
\pgfpathlineto{\pgfqpoint{3.180044in}{0.341797in}}%
\pgfpathlineto{\pgfqpoint{3.259727in}{0.341307in}}%
\pgfpathlineto{\pgfqpoint{3.339410in}{0.337140in}}%
\pgfpathlineto{\pgfqpoint{3.697984in}{0.310040in}}%
\pgfpathlineto{\pgfqpoint{3.857350in}{0.295098in}}%
\pgfpathlineto{\pgfqpoint{3.937034in}{0.288708in}}%
\pgfpathlineto{\pgfqpoint{3.976875in}{0.287290in}}%
\pgfpathlineto{\pgfqpoint{4.016717in}{0.288111in}}%
\pgfpathlineto{\pgfqpoint{4.056558in}{0.292180in}}%
\pgfpathlineto{\pgfqpoint{4.096400in}{0.300576in}}%
\pgfpathlineto{\pgfqpoint{4.136241in}{0.314286in}}%
\pgfpathlineto{\pgfqpoint{4.176083in}{0.333975in}}%
\pgfpathlineto{\pgfqpoint{4.215924in}{0.359715in}}%
\pgfpathlineto{\pgfqpoint{4.255766in}{0.390750in}}%
\pgfpathlineto{\pgfqpoint{4.375291in}{0.493841in}}%
\pgfpathlineto{\pgfqpoint{4.415132in}{0.520739in}}%
\pgfpathlineto{\pgfqpoint{4.454974in}{0.538233in}}%
\pgfpathlineto{\pgfqpoint{4.494815in}{0.543939in}}%
\pgfpathlineto{\pgfqpoint{4.510000in}{0.541229in}}%
\pgfpathlineto{\pgfqpoint{4.510000in}{0.541229in}}%
\pgfusepath{stroke}%
\end{pgfscope}%
\begin{pgfscope}%
\pgfpathrectangle{\pgfqpoint{0.625000in}{0.200000in}}{\pgfqpoint{3.875000in}{1.600000in}} %
\pgfusepath{clip}%
\pgfsetbuttcap%
\pgfsetroundjoin%
\pgfsetlinewidth{2.007500pt}%
\definecolor{currentstroke}{rgb}{0.400000,0.400000,0.400000}%
\pgfsetstrokecolor{currentstroke}%
\pgfsetdash{{1.000000pt}{3.000000pt}}{0.000000pt}%
\pgfpathmoveto{\pgfqpoint{0.615000in}{0.262742in}}%
\pgfpathlineto{\pgfqpoint{0.636710in}{0.264502in}}%
\pgfpathlineto{\pgfqpoint{0.672706in}{0.262656in}}%
\pgfpathlineto{\pgfqpoint{0.708702in}{0.257231in}}%
\pgfpathlineto{\pgfqpoint{0.780694in}{0.243961in}}%
\pgfpathlineto{\pgfqpoint{0.816690in}{0.239731in}}%
\pgfpathlineto{\pgfqpoint{0.852687in}{0.238154in}}%
\pgfpathlineto{\pgfqpoint{0.888683in}{0.238991in}}%
\pgfpathlineto{\pgfqpoint{0.960675in}{0.245173in}}%
\pgfpathlineto{\pgfqpoint{1.140655in}{0.264283in}}%
\pgfpathlineto{\pgfqpoint{1.320635in}{0.281333in}}%
\pgfpathlineto{\pgfqpoint{1.500616in}{0.301147in}}%
\pgfpathlineto{\pgfqpoint{1.572608in}{0.309016in}}%
\pgfpathlineto{\pgfqpoint{1.644600in}{0.314286in}}%
\pgfpathlineto{\pgfqpoint{1.788584in}{0.321789in}}%
\pgfpathlineto{\pgfqpoint{1.860577in}{0.329108in}}%
\pgfpathlineto{\pgfqpoint{2.004561in}{0.347214in}}%
\pgfpathlineto{\pgfqpoint{2.112549in}{0.364638in}}%
\pgfpathlineto{\pgfqpoint{2.184541in}{0.375066in}}%
\pgfpathlineto{\pgfqpoint{2.256533in}{0.381230in}}%
\pgfpathlineto{\pgfqpoint{2.328526in}{0.385527in}}%
\pgfpathlineto{\pgfqpoint{2.364522in}{0.389191in}}%
\pgfpathlineto{\pgfqpoint{2.400518in}{0.395098in}}%
\pgfpathlineto{\pgfqpoint{2.436514in}{0.403443in}}%
\pgfpathlineto{\pgfqpoint{2.508506in}{0.423002in}}%
\pgfpathlineto{\pgfqpoint{2.544502in}{0.430300in}}%
\pgfpathlineto{\pgfqpoint{2.580498in}{0.433963in}}%
\pgfpathlineto{\pgfqpoint{2.616494in}{0.434239in}}%
\pgfpathlineto{\pgfqpoint{2.688486in}{0.432087in}}%
\pgfpathlineto{\pgfqpoint{2.724482in}{0.433822in}}%
\pgfpathlineto{\pgfqpoint{2.760478in}{0.438697in}}%
\pgfpathlineto{\pgfqpoint{2.796474in}{0.446163in}}%
\pgfpathlineto{\pgfqpoint{2.904463in}{0.471666in}}%
\pgfpathlineto{\pgfqpoint{2.976455in}{0.483726in}}%
\pgfpathlineto{\pgfqpoint{3.048447in}{0.494070in}}%
\pgfpathlineto{\pgfqpoint{3.120439in}{0.507541in}}%
\pgfpathlineto{\pgfqpoint{3.300419in}{0.547256in}}%
\pgfpathlineto{\pgfqpoint{3.336416in}{0.557245in}}%
\pgfpathlineto{\pgfqpoint{3.372412in}{0.569063in}}%
\pgfpathlineto{\pgfqpoint{3.480400in}{0.608769in}}%
\pgfpathlineto{\pgfqpoint{3.516396in}{0.619801in}}%
\pgfpathlineto{\pgfqpoint{3.552392in}{0.628754in}}%
\pgfpathlineto{\pgfqpoint{3.624384in}{0.642369in}}%
\pgfpathlineto{\pgfqpoint{3.804365in}{0.672569in}}%
\pgfpathlineto{\pgfqpoint{3.876357in}{0.681512in}}%
\pgfpathlineto{\pgfqpoint{3.912353in}{0.683863in}}%
\pgfpathlineto{\pgfqpoint{3.948349in}{0.683687in}}%
\pgfpathlineto{\pgfqpoint{3.984345in}{0.680012in}}%
\pgfpathlineto{\pgfqpoint{4.020341in}{0.672015in}}%
\pgfpathlineto{\pgfqpoint{4.056337in}{0.659160in}}%
\pgfpathlineto{\pgfqpoint{4.092333in}{0.641317in}}%
\pgfpathlineto{\pgfqpoint{4.128329in}{0.618976in}}%
\pgfpathlineto{\pgfqpoint{4.200321in}{0.568669in}}%
\pgfpathlineto{\pgfqpoint{4.236317in}{0.549191in}}%
\pgfpathlineto{\pgfqpoint{4.272313in}{0.542133in}}%
\pgfpathlineto{\pgfqpoint{4.308310in}{0.554098in}}%
\pgfpathlineto{\pgfqpoint{4.344306in}{0.588176in}}%
\pgfpathlineto{\pgfqpoint{4.380302in}{0.640530in}}%
\pgfpathlineto{\pgfqpoint{4.416298in}{0.698953in}}%
\pgfpathlineto{\pgfqpoint{4.452294in}{0.745222in}}%
\pgfpathlineto{\pgfqpoint{4.488290in}{0.761231in}}%
\pgfpathlineto{\pgfqpoint{4.510000in}{0.746195in}}%
\pgfpathlineto{\pgfqpoint{4.510000in}{0.746195in}}%
\pgfusepath{stroke}%
\end{pgfscope}%
\begin{pgfscope}%
\pgfsetrectcap%
\pgfsetmiterjoin%
\pgfsetlinewidth{1.003750pt}%
\definecolor{currentstroke}{rgb}{0.000000,0.000000,0.000000}%
\pgfsetstrokecolor{currentstroke}%
\pgfsetdash{}{0pt}%
\pgfpathmoveto{\pgfqpoint{0.625000in}{1.800000in}}%
\pgfpathlineto{\pgfqpoint{4.500000in}{1.800000in}}%
\pgfusepath{stroke}%
\end{pgfscope}%
\begin{pgfscope}%
\pgfsetrectcap%
\pgfsetmiterjoin%
\pgfsetlinewidth{1.003750pt}%
\definecolor{currentstroke}{rgb}{0.000000,0.000000,0.000000}%
\pgfsetstrokecolor{currentstroke}%
\pgfsetdash{}{0pt}%
\pgfpathmoveto{\pgfqpoint{0.625000in}{0.200000in}}%
\pgfpathlineto{\pgfqpoint{4.500000in}{0.200000in}}%
\pgfusepath{stroke}%
\end{pgfscope}%
\begin{pgfscope}%
\pgfsetrectcap%
\pgfsetmiterjoin%
\pgfsetlinewidth{1.003750pt}%
\definecolor{currentstroke}{rgb}{0.000000,0.000000,0.000000}%
\pgfsetstrokecolor{currentstroke}%
\pgfsetdash{}{0pt}%
\pgfpathmoveto{\pgfqpoint{4.500000in}{0.200000in}}%
\pgfpathlineto{\pgfqpoint{4.500000in}{1.800000in}}%
\pgfusepath{stroke}%
\end{pgfscope}%
\begin{pgfscope}%
\pgfsetrectcap%
\pgfsetmiterjoin%
\pgfsetlinewidth{1.003750pt}%
\definecolor{currentstroke}{rgb}{0.000000,0.000000,0.000000}%
\pgfsetstrokecolor{currentstroke}%
\pgfsetdash{}{0pt}%
\pgfpathmoveto{\pgfqpoint{0.625000in}{0.200000in}}%
\pgfpathlineto{\pgfqpoint{0.625000in}{1.800000in}}%
\pgfusepath{stroke}%
\end{pgfscope}%
\begin{pgfscope}%
\pgfsetbuttcap%
\pgfsetroundjoin%
\definecolor{currentfill}{rgb}{0.000000,0.000000,0.000000}%
\pgfsetfillcolor{currentfill}%
\pgfsetlinewidth{0.501875pt}%
\definecolor{currentstroke}{rgb}{0.000000,0.000000,0.000000}%
\pgfsetstrokecolor{currentstroke}%
\pgfsetdash{}{0pt}%
\pgfsys@defobject{currentmarker}{\pgfqpoint{0.000000in}{0.000000in}}{\pgfqpoint{0.000000in}{0.055556in}}{%
\pgfpathmoveto{\pgfqpoint{0.000000in}{0.000000in}}%
\pgfpathlineto{\pgfqpoint{0.000000in}{0.055556in}}%
\pgfusepath{stroke,fill}%
}%
\begin{pgfscope}%
\pgfsys@transformshift{0.625000in}{0.200000in}%
\pgfsys@useobject{currentmarker}{}%
\end{pgfscope}%
\end{pgfscope}%
\begin{pgfscope}%
\pgfsetbuttcap%
\pgfsetroundjoin%
\definecolor{currentfill}{rgb}{0.000000,0.000000,0.000000}%
\pgfsetfillcolor{currentfill}%
\pgfsetlinewidth{0.501875pt}%
\definecolor{currentstroke}{rgb}{0.000000,0.000000,0.000000}%
\pgfsetstrokecolor{currentstroke}%
\pgfsetdash{}{0pt}%
\pgfsys@defobject{currentmarker}{\pgfqpoint{0.000000in}{-0.055556in}}{\pgfqpoint{0.000000in}{0.000000in}}{%
\pgfpathmoveto{\pgfqpoint{0.000000in}{0.000000in}}%
\pgfpathlineto{\pgfqpoint{0.000000in}{-0.055556in}}%
\pgfusepath{stroke,fill}%
}%
\begin{pgfscope}%
\pgfsys@transformshift{0.625000in}{1.800000in}%
\pgfsys@useobject{currentmarker}{}%
\end{pgfscope}%
\end{pgfscope}%
\begin{pgfscope}%
\pgftext[x=0.625000in,y=0.144444in,,top]{\sffamily\fontsize{12.000000}{14.400000}\selectfont 0.0}%
\end{pgfscope}%
\begin{pgfscope}%
\pgfsetbuttcap%
\pgfsetroundjoin%
\definecolor{currentfill}{rgb}{0.000000,0.000000,0.000000}%
\pgfsetfillcolor{currentfill}%
\pgfsetlinewidth{0.501875pt}%
\definecolor{currentstroke}{rgb}{0.000000,0.000000,0.000000}%
\pgfsetstrokecolor{currentstroke}%
\pgfsetdash{}{0pt}%
\pgfsys@defobject{currentmarker}{\pgfqpoint{0.000000in}{0.000000in}}{\pgfqpoint{0.000000in}{0.055556in}}{%
\pgfpathmoveto{\pgfqpoint{0.000000in}{0.000000in}}%
\pgfpathlineto{\pgfqpoint{0.000000in}{0.055556in}}%
\pgfusepath{stroke,fill}%
}%
\begin{pgfscope}%
\pgfsys@transformshift{1.400000in}{0.200000in}%
\pgfsys@useobject{currentmarker}{}%
\end{pgfscope}%
\end{pgfscope}%
\begin{pgfscope}%
\pgfsetbuttcap%
\pgfsetroundjoin%
\definecolor{currentfill}{rgb}{0.000000,0.000000,0.000000}%
\pgfsetfillcolor{currentfill}%
\pgfsetlinewidth{0.501875pt}%
\definecolor{currentstroke}{rgb}{0.000000,0.000000,0.000000}%
\pgfsetstrokecolor{currentstroke}%
\pgfsetdash{}{0pt}%
\pgfsys@defobject{currentmarker}{\pgfqpoint{0.000000in}{-0.055556in}}{\pgfqpoint{0.000000in}{0.000000in}}{%
\pgfpathmoveto{\pgfqpoint{0.000000in}{0.000000in}}%
\pgfpathlineto{\pgfqpoint{0.000000in}{-0.055556in}}%
\pgfusepath{stroke,fill}%
}%
\begin{pgfscope}%
\pgfsys@transformshift{1.400000in}{1.800000in}%
\pgfsys@useobject{currentmarker}{}%
\end{pgfscope}%
\end{pgfscope}%
\begin{pgfscope}%
\pgftext[x=1.400000in,y=0.144444in,,top]{\sffamily\fontsize{12.000000}{14.400000}\selectfont 0.2}%
\end{pgfscope}%
\begin{pgfscope}%
\pgfsetbuttcap%
\pgfsetroundjoin%
\definecolor{currentfill}{rgb}{0.000000,0.000000,0.000000}%
\pgfsetfillcolor{currentfill}%
\pgfsetlinewidth{0.501875pt}%
\definecolor{currentstroke}{rgb}{0.000000,0.000000,0.000000}%
\pgfsetstrokecolor{currentstroke}%
\pgfsetdash{}{0pt}%
\pgfsys@defobject{currentmarker}{\pgfqpoint{0.000000in}{0.000000in}}{\pgfqpoint{0.000000in}{0.055556in}}{%
\pgfpathmoveto{\pgfqpoint{0.000000in}{0.000000in}}%
\pgfpathlineto{\pgfqpoint{0.000000in}{0.055556in}}%
\pgfusepath{stroke,fill}%
}%
\begin{pgfscope}%
\pgfsys@transformshift{2.175000in}{0.200000in}%
\pgfsys@useobject{currentmarker}{}%
\end{pgfscope}%
\end{pgfscope}%
\begin{pgfscope}%
\pgfsetbuttcap%
\pgfsetroundjoin%
\definecolor{currentfill}{rgb}{0.000000,0.000000,0.000000}%
\pgfsetfillcolor{currentfill}%
\pgfsetlinewidth{0.501875pt}%
\definecolor{currentstroke}{rgb}{0.000000,0.000000,0.000000}%
\pgfsetstrokecolor{currentstroke}%
\pgfsetdash{}{0pt}%
\pgfsys@defobject{currentmarker}{\pgfqpoint{0.000000in}{-0.055556in}}{\pgfqpoint{0.000000in}{0.000000in}}{%
\pgfpathmoveto{\pgfqpoint{0.000000in}{0.000000in}}%
\pgfpathlineto{\pgfqpoint{0.000000in}{-0.055556in}}%
\pgfusepath{stroke,fill}%
}%
\begin{pgfscope}%
\pgfsys@transformshift{2.175000in}{1.800000in}%
\pgfsys@useobject{currentmarker}{}%
\end{pgfscope}%
\end{pgfscope}%
\begin{pgfscope}%
\pgftext[x=2.175000in,y=0.144444in,,top]{\sffamily\fontsize{12.000000}{14.400000}\selectfont 0.4}%
\end{pgfscope}%
\begin{pgfscope}%
\pgfsetbuttcap%
\pgfsetroundjoin%
\definecolor{currentfill}{rgb}{0.000000,0.000000,0.000000}%
\pgfsetfillcolor{currentfill}%
\pgfsetlinewidth{0.501875pt}%
\definecolor{currentstroke}{rgb}{0.000000,0.000000,0.000000}%
\pgfsetstrokecolor{currentstroke}%
\pgfsetdash{}{0pt}%
\pgfsys@defobject{currentmarker}{\pgfqpoint{0.000000in}{0.000000in}}{\pgfqpoint{0.000000in}{0.055556in}}{%
\pgfpathmoveto{\pgfqpoint{0.000000in}{0.000000in}}%
\pgfpathlineto{\pgfqpoint{0.000000in}{0.055556in}}%
\pgfusepath{stroke,fill}%
}%
\begin{pgfscope}%
\pgfsys@transformshift{2.950000in}{0.200000in}%
\pgfsys@useobject{currentmarker}{}%
\end{pgfscope}%
\end{pgfscope}%
\begin{pgfscope}%
\pgfsetbuttcap%
\pgfsetroundjoin%
\definecolor{currentfill}{rgb}{0.000000,0.000000,0.000000}%
\pgfsetfillcolor{currentfill}%
\pgfsetlinewidth{0.501875pt}%
\definecolor{currentstroke}{rgb}{0.000000,0.000000,0.000000}%
\pgfsetstrokecolor{currentstroke}%
\pgfsetdash{}{0pt}%
\pgfsys@defobject{currentmarker}{\pgfqpoint{0.000000in}{-0.055556in}}{\pgfqpoint{0.000000in}{0.000000in}}{%
\pgfpathmoveto{\pgfqpoint{0.000000in}{0.000000in}}%
\pgfpathlineto{\pgfqpoint{0.000000in}{-0.055556in}}%
\pgfusepath{stroke,fill}%
}%
\begin{pgfscope}%
\pgfsys@transformshift{2.950000in}{1.800000in}%
\pgfsys@useobject{currentmarker}{}%
\end{pgfscope}%
\end{pgfscope}%
\begin{pgfscope}%
\pgftext[x=2.950000in,y=0.144444in,,top]{\sffamily\fontsize{12.000000}{14.400000}\selectfont 0.6}%
\end{pgfscope}%
\begin{pgfscope}%
\pgfsetbuttcap%
\pgfsetroundjoin%
\definecolor{currentfill}{rgb}{0.000000,0.000000,0.000000}%
\pgfsetfillcolor{currentfill}%
\pgfsetlinewidth{0.501875pt}%
\definecolor{currentstroke}{rgb}{0.000000,0.000000,0.000000}%
\pgfsetstrokecolor{currentstroke}%
\pgfsetdash{}{0pt}%
\pgfsys@defobject{currentmarker}{\pgfqpoint{0.000000in}{0.000000in}}{\pgfqpoint{0.000000in}{0.055556in}}{%
\pgfpathmoveto{\pgfqpoint{0.000000in}{0.000000in}}%
\pgfpathlineto{\pgfqpoint{0.000000in}{0.055556in}}%
\pgfusepath{stroke,fill}%
}%
\begin{pgfscope}%
\pgfsys@transformshift{3.725000in}{0.200000in}%
\pgfsys@useobject{currentmarker}{}%
\end{pgfscope}%
\end{pgfscope}%
\begin{pgfscope}%
\pgfsetbuttcap%
\pgfsetroundjoin%
\definecolor{currentfill}{rgb}{0.000000,0.000000,0.000000}%
\pgfsetfillcolor{currentfill}%
\pgfsetlinewidth{0.501875pt}%
\definecolor{currentstroke}{rgb}{0.000000,0.000000,0.000000}%
\pgfsetstrokecolor{currentstroke}%
\pgfsetdash{}{0pt}%
\pgfsys@defobject{currentmarker}{\pgfqpoint{0.000000in}{-0.055556in}}{\pgfqpoint{0.000000in}{0.000000in}}{%
\pgfpathmoveto{\pgfqpoint{0.000000in}{0.000000in}}%
\pgfpathlineto{\pgfqpoint{0.000000in}{-0.055556in}}%
\pgfusepath{stroke,fill}%
}%
\begin{pgfscope}%
\pgfsys@transformshift{3.725000in}{1.800000in}%
\pgfsys@useobject{currentmarker}{}%
\end{pgfscope}%
\end{pgfscope}%
\begin{pgfscope}%
\pgftext[x=3.725000in,y=0.144444in,,top]{\sffamily\fontsize{12.000000}{14.400000}\selectfont 0.8}%
\end{pgfscope}%
\begin{pgfscope}%
\pgfsetbuttcap%
\pgfsetroundjoin%
\definecolor{currentfill}{rgb}{0.000000,0.000000,0.000000}%
\pgfsetfillcolor{currentfill}%
\pgfsetlinewidth{0.501875pt}%
\definecolor{currentstroke}{rgb}{0.000000,0.000000,0.000000}%
\pgfsetstrokecolor{currentstroke}%
\pgfsetdash{}{0pt}%
\pgfsys@defobject{currentmarker}{\pgfqpoint{0.000000in}{0.000000in}}{\pgfqpoint{0.000000in}{0.055556in}}{%
\pgfpathmoveto{\pgfqpoint{0.000000in}{0.000000in}}%
\pgfpathlineto{\pgfqpoint{0.000000in}{0.055556in}}%
\pgfusepath{stroke,fill}%
}%
\begin{pgfscope}%
\pgfsys@transformshift{4.500000in}{0.200000in}%
\pgfsys@useobject{currentmarker}{}%
\end{pgfscope}%
\end{pgfscope}%
\begin{pgfscope}%
\pgfsetbuttcap%
\pgfsetroundjoin%
\definecolor{currentfill}{rgb}{0.000000,0.000000,0.000000}%
\pgfsetfillcolor{currentfill}%
\pgfsetlinewidth{0.501875pt}%
\definecolor{currentstroke}{rgb}{0.000000,0.000000,0.000000}%
\pgfsetstrokecolor{currentstroke}%
\pgfsetdash{}{0pt}%
\pgfsys@defobject{currentmarker}{\pgfqpoint{0.000000in}{-0.055556in}}{\pgfqpoint{0.000000in}{0.000000in}}{%
\pgfpathmoveto{\pgfqpoint{0.000000in}{0.000000in}}%
\pgfpathlineto{\pgfqpoint{0.000000in}{-0.055556in}}%
\pgfusepath{stroke,fill}%
}%
\begin{pgfscope}%
\pgfsys@transformshift{4.500000in}{1.800000in}%
\pgfsys@useobject{currentmarker}{}%
\end{pgfscope}%
\end{pgfscope}%
\begin{pgfscope}%
\pgftext[x=4.500000in,y=0.144444in,,top]{\sffamily\fontsize{12.000000}{14.400000}\selectfont 1.0}%
\end{pgfscope}%
\begin{pgfscope}%
\pgfsetbuttcap%
\pgfsetroundjoin%
\definecolor{currentfill}{rgb}{0.000000,0.000000,0.000000}%
\pgfsetfillcolor{currentfill}%
\pgfsetlinewidth{0.501875pt}%
\definecolor{currentstroke}{rgb}{0.000000,0.000000,0.000000}%
\pgfsetstrokecolor{currentstroke}%
\pgfsetdash{}{0pt}%
\pgfsys@defobject{currentmarker}{\pgfqpoint{0.000000in}{0.000000in}}{\pgfqpoint{0.055556in}{0.000000in}}{%
\pgfpathmoveto{\pgfqpoint{0.000000in}{0.000000in}}%
\pgfpathlineto{\pgfqpoint{0.055556in}{0.000000in}}%
\pgfusepath{stroke,fill}%
}%
\begin{pgfscope}%
\pgfsys@transformshift{0.625000in}{0.200000in}%
\pgfsys@useobject{currentmarker}{}%
\end{pgfscope}%
\end{pgfscope}%
\begin{pgfscope}%
\pgfsetbuttcap%
\pgfsetroundjoin%
\definecolor{currentfill}{rgb}{0.000000,0.000000,0.000000}%
\pgfsetfillcolor{currentfill}%
\pgfsetlinewidth{0.501875pt}%
\definecolor{currentstroke}{rgb}{0.000000,0.000000,0.000000}%
\pgfsetstrokecolor{currentstroke}%
\pgfsetdash{}{0pt}%
\pgfsys@defobject{currentmarker}{\pgfqpoint{-0.055556in}{0.000000in}}{\pgfqpoint{0.000000in}{0.000000in}}{%
\pgfpathmoveto{\pgfqpoint{0.000000in}{0.000000in}}%
\pgfpathlineto{\pgfqpoint{-0.055556in}{0.000000in}}%
\pgfusepath{stroke,fill}%
}%
\begin{pgfscope}%
\pgfsys@transformshift{4.500000in}{0.200000in}%
\pgfsys@useobject{currentmarker}{}%
\end{pgfscope}%
\end{pgfscope}%
\begin{pgfscope}%
\pgftext[x=0.569444in,y=0.200000in,right,]{\sffamily\fontsize{12.000000}{14.400000}\selectfont 0}%
\end{pgfscope}%
\begin{pgfscope}%
\pgfsetbuttcap%
\pgfsetroundjoin%
\definecolor{currentfill}{rgb}{0.000000,0.000000,0.000000}%
\pgfsetfillcolor{currentfill}%
\pgfsetlinewidth{0.501875pt}%
\definecolor{currentstroke}{rgb}{0.000000,0.000000,0.000000}%
\pgfsetstrokecolor{currentstroke}%
\pgfsetdash{}{0pt}%
\pgfsys@defobject{currentmarker}{\pgfqpoint{0.000000in}{0.000000in}}{\pgfqpoint{0.055556in}{0.000000in}}{%
\pgfpathmoveto{\pgfqpoint{0.000000in}{0.000000in}}%
\pgfpathlineto{\pgfqpoint{0.055556in}{0.000000in}}%
\pgfusepath{stroke,fill}%
}%
\begin{pgfscope}%
\pgfsys@transformshift{0.625000in}{0.466667in}%
\pgfsys@useobject{currentmarker}{}%
\end{pgfscope}%
\end{pgfscope}%
\begin{pgfscope}%
\pgfsetbuttcap%
\pgfsetroundjoin%
\definecolor{currentfill}{rgb}{0.000000,0.000000,0.000000}%
\pgfsetfillcolor{currentfill}%
\pgfsetlinewidth{0.501875pt}%
\definecolor{currentstroke}{rgb}{0.000000,0.000000,0.000000}%
\pgfsetstrokecolor{currentstroke}%
\pgfsetdash{}{0pt}%
\pgfsys@defobject{currentmarker}{\pgfqpoint{-0.055556in}{0.000000in}}{\pgfqpoint{0.000000in}{0.000000in}}{%
\pgfpathmoveto{\pgfqpoint{0.000000in}{0.000000in}}%
\pgfpathlineto{\pgfqpoint{-0.055556in}{0.000000in}}%
\pgfusepath{stroke,fill}%
}%
\begin{pgfscope}%
\pgfsys@transformshift{4.500000in}{0.466667in}%
\pgfsys@useobject{currentmarker}{}%
\end{pgfscope}%
\end{pgfscope}%
\begin{pgfscope}%
\pgftext[x=0.569444in,y=0.466667in,right,]{\sffamily\fontsize{12.000000}{14.400000}\selectfont 1}%
\end{pgfscope}%
\begin{pgfscope}%
\pgfsetbuttcap%
\pgfsetroundjoin%
\definecolor{currentfill}{rgb}{0.000000,0.000000,0.000000}%
\pgfsetfillcolor{currentfill}%
\pgfsetlinewidth{0.501875pt}%
\definecolor{currentstroke}{rgb}{0.000000,0.000000,0.000000}%
\pgfsetstrokecolor{currentstroke}%
\pgfsetdash{}{0pt}%
\pgfsys@defobject{currentmarker}{\pgfqpoint{0.000000in}{0.000000in}}{\pgfqpoint{0.055556in}{0.000000in}}{%
\pgfpathmoveto{\pgfqpoint{0.000000in}{0.000000in}}%
\pgfpathlineto{\pgfqpoint{0.055556in}{0.000000in}}%
\pgfusepath{stroke,fill}%
}%
\begin{pgfscope}%
\pgfsys@transformshift{0.625000in}{0.733333in}%
\pgfsys@useobject{currentmarker}{}%
\end{pgfscope}%
\end{pgfscope}%
\begin{pgfscope}%
\pgfsetbuttcap%
\pgfsetroundjoin%
\definecolor{currentfill}{rgb}{0.000000,0.000000,0.000000}%
\pgfsetfillcolor{currentfill}%
\pgfsetlinewidth{0.501875pt}%
\definecolor{currentstroke}{rgb}{0.000000,0.000000,0.000000}%
\pgfsetstrokecolor{currentstroke}%
\pgfsetdash{}{0pt}%
\pgfsys@defobject{currentmarker}{\pgfqpoint{-0.055556in}{0.000000in}}{\pgfqpoint{0.000000in}{0.000000in}}{%
\pgfpathmoveto{\pgfqpoint{0.000000in}{0.000000in}}%
\pgfpathlineto{\pgfqpoint{-0.055556in}{0.000000in}}%
\pgfusepath{stroke,fill}%
}%
\begin{pgfscope}%
\pgfsys@transformshift{4.500000in}{0.733333in}%
\pgfsys@useobject{currentmarker}{}%
\end{pgfscope}%
\end{pgfscope}%
\begin{pgfscope}%
\pgftext[x=0.569444in,y=0.733333in,right,]{\sffamily\fontsize{12.000000}{14.400000}\selectfont 2}%
\end{pgfscope}%
\begin{pgfscope}%
\pgfsetbuttcap%
\pgfsetroundjoin%
\definecolor{currentfill}{rgb}{0.000000,0.000000,0.000000}%
\pgfsetfillcolor{currentfill}%
\pgfsetlinewidth{0.501875pt}%
\definecolor{currentstroke}{rgb}{0.000000,0.000000,0.000000}%
\pgfsetstrokecolor{currentstroke}%
\pgfsetdash{}{0pt}%
\pgfsys@defobject{currentmarker}{\pgfqpoint{0.000000in}{0.000000in}}{\pgfqpoint{0.055556in}{0.000000in}}{%
\pgfpathmoveto{\pgfqpoint{0.000000in}{0.000000in}}%
\pgfpathlineto{\pgfqpoint{0.055556in}{0.000000in}}%
\pgfusepath{stroke,fill}%
}%
\begin{pgfscope}%
\pgfsys@transformshift{0.625000in}{1.000000in}%
\pgfsys@useobject{currentmarker}{}%
\end{pgfscope}%
\end{pgfscope}%
\begin{pgfscope}%
\pgfsetbuttcap%
\pgfsetroundjoin%
\definecolor{currentfill}{rgb}{0.000000,0.000000,0.000000}%
\pgfsetfillcolor{currentfill}%
\pgfsetlinewidth{0.501875pt}%
\definecolor{currentstroke}{rgb}{0.000000,0.000000,0.000000}%
\pgfsetstrokecolor{currentstroke}%
\pgfsetdash{}{0pt}%
\pgfsys@defobject{currentmarker}{\pgfqpoint{-0.055556in}{0.000000in}}{\pgfqpoint{0.000000in}{0.000000in}}{%
\pgfpathmoveto{\pgfqpoint{0.000000in}{0.000000in}}%
\pgfpathlineto{\pgfqpoint{-0.055556in}{0.000000in}}%
\pgfusepath{stroke,fill}%
}%
\begin{pgfscope}%
\pgfsys@transformshift{4.500000in}{1.000000in}%
\pgfsys@useobject{currentmarker}{}%
\end{pgfscope}%
\end{pgfscope}%
\begin{pgfscope}%
\pgftext[x=0.569444in,y=1.000000in,right,]{\sffamily\fontsize{12.000000}{14.400000}\selectfont 3}%
\end{pgfscope}%
\begin{pgfscope}%
\pgfsetbuttcap%
\pgfsetroundjoin%
\definecolor{currentfill}{rgb}{0.000000,0.000000,0.000000}%
\pgfsetfillcolor{currentfill}%
\pgfsetlinewidth{0.501875pt}%
\definecolor{currentstroke}{rgb}{0.000000,0.000000,0.000000}%
\pgfsetstrokecolor{currentstroke}%
\pgfsetdash{}{0pt}%
\pgfsys@defobject{currentmarker}{\pgfqpoint{0.000000in}{0.000000in}}{\pgfqpoint{0.055556in}{0.000000in}}{%
\pgfpathmoveto{\pgfqpoint{0.000000in}{0.000000in}}%
\pgfpathlineto{\pgfqpoint{0.055556in}{0.000000in}}%
\pgfusepath{stroke,fill}%
}%
\begin{pgfscope}%
\pgfsys@transformshift{0.625000in}{1.266667in}%
\pgfsys@useobject{currentmarker}{}%
\end{pgfscope}%
\end{pgfscope}%
\begin{pgfscope}%
\pgfsetbuttcap%
\pgfsetroundjoin%
\definecolor{currentfill}{rgb}{0.000000,0.000000,0.000000}%
\pgfsetfillcolor{currentfill}%
\pgfsetlinewidth{0.501875pt}%
\definecolor{currentstroke}{rgb}{0.000000,0.000000,0.000000}%
\pgfsetstrokecolor{currentstroke}%
\pgfsetdash{}{0pt}%
\pgfsys@defobject{currentmarker}{\pgfqpoint{-0.055556in}{0.000000in}}{\pgfqpoint{0.000000in}{0.000000in}}{%
\pgfpathmoveto{\pgfqpoint{0.000000in}{0.000000in}}%
\pgfpathlineto{\pgfqpoint{-0.055556in}{0.000000in}}%
\pgfusepath{stroke,fill}%
}%
\begin{pgfscope}%
\pgfsys@transformshift{4.500000in}{1.266667in}%
\pgfsys@useobject{currentmarker}{}%
\end{pgfscope}%
\end{pgfscope}%
\begin{pgfscope}%
\pgftext[x=0.569444in,y=1.266667in,right,]{\sffamily\fontsize{12.000000}{14.400000}\selectfont 4}%
\end{pgfscope}%
\begin{pgfscope}%
\pgfsetbuttcap%
\pgfsetroundjoin%
\definecolor{currentfill}{rgb}{0.000000,0.000000,0.000000}%
\pgfsetfillcolor{currentfill}%
\pgfsetlinewidth{0.501875pt}%
\definecolor{currentstroke}{rgb}{0.000000,0.000000,0.000000}%
\pgfsetstrokecolor{currentstroke}%
\pgfsetdash{}{0pt}%
\pgfsys@defobject{currentmarker}{\pgfqpoint{0.000000in}{0.000000in}}{\pgfqpoint{0.055556in}{0.000000in}}{%
\pgfpathmoveto{\pgfqpoint{0.000000in}{0.000000in}}%
\pgfpathlineto{\pgfqpoint{0.055556in}{0.000000in}}%
\pgfusepath{stroke,fill}%
}%
\begin{pgfscope}%
\pgfsys@transformshift{0.625000in}{1.533333in}%
\pgfsys@useobject{currentmarker}{}%
\end{pgfscope}%
\end{pgfscope}%
\begin{pgfscope}%
\pgfsetbuttcap%
\pgfsetroundjoin%
\definecolor{currentfill}{rgb}{0.000000,0.000000,0.000000}%
\pgfsetfillcolor{currentfill}%
\pgfsetlinewidth{0.501875pt}%
\definecolor{currentstroke}{rgb}{0.000000,0.000000,0.000000}%
\pgfsetstrokecolor{currentstroke}%
\pgfsetdash{}{0pt}%
\pgfsys@defobject{currentmarker}{\pgfqpoint{-0.055556in}{0.000000in}}{\pgfqpoint{0.000000in}{0.000000in}}{%
\pgfpathmoveto{\pgfqpoint{0.000000in}{0.000000in}}%
\pgfpathlineto{\pgfqpoint{-0.055556in}{0.000000in}}%
\pgfusepath{stroke,fill}%
}%
\begin{pgfscope}%
\pgfsys@transformshift{4.500000in}{1.533333in}%
\pgfsys@useobject{currentmarker}{}%
\end{pgfscope}%
\end{pgfscope}%
\begin{pgfscope}%
\pgftext[x=0.569444in,y=1.533333in,right,]{\sffamily\fontsize{12.000000}{14.400000}\selectfont 5}%
\end{pgfscope}%
\begin{pgfscope}%
\pgfsetbuttcap%
\pgfsetroundjoin%
\definecolor{currentfill}{rgb}{0.000000,0.000000,0.000000}%
\pgfsetfillcolor{currentfill}%
\pgfsetlinewidth{0.501875pt}%
\definecolor{currentstroke}{rgb}{0.000000,0.000000,0.000000}%
\pgfsetstrokecolor{currentstroke}%
\pgfsetdash{}{0pt}%
\pgfsys@defobject{currentmarker}{\pgfqpoint{0.000000in}{0.000000in}}{\pgfqpoint{0.055556in}{0.000000in}}{%
\pgfpathmoveto{\pgfqpoint{0.000000in}{0.000000in}}%
\pgfpathlineto{\pgfqpoint{0.055556in}{0.000000in}}%
\pgfusepath{stroke,fill}%
}%
\begin{pgfscope}%
\pgfsys@transformshift{0.625000in}{1.800000in}%
\pgfsys@useobject{currentmarker}{}%
\end{pgfscope}%
\end{pgfscope}%
\begin{pgfscope}%
\pgfsetbuttcap%
\pgfsetroundjoin%
\definecolor{currentfill}{rgb}{0.000000,0.000000,0.000000}%
\pgfsetfillcolor{currentfill}%
\pgfsetlinewidth{0.501875pt}%
\definecolor{currentstroke}{rgb}{0.000000,0.000000,0.000000}%
\pgfsetstrokecolor{currentstroke}%
\pgfsetdash{}{0pt}%
\pgfsys@defobject{currentmarker}{\pgfqpoint{-0.055556in}{0.000000in}}{\pgfqpoint{0.000000in}{0.000000in}}{%
\pgfpathmoveto{\pgfqpoint{0.000000in}{0.000000in}}%
\pgfpathlineto{\pgfqpoint{-0.055556in}{0.000000in}}%
\pgfusepath{stroke,fill}%
}%
\begin{pgfscope}%
\pgfsys@transformshift{4.500000in}{1.800000in}%
\pgfsys@useobject{currentmarker}{}%
\end{pgfscope}%
\end{pgfscope}%
\begin{pgfscope}%
\pgftext[x=0.569444in,y=1.800000in,right,]{\sffamily\fontsize{12.000000}{14.400000}\selectfont 6}%
\end{pgfscope}%
\begin{pgfscope}%
\pgftext[x=0.394613in,y=1.000000in,,bottom,rotate=90.000000]{\sffamily\fontsize{12.000000}{14.400000}\selectfont Density}%
\end{pgfscope}%
\begin{pgfscope}%
\pgfsetbuttcap%
\pgfsetmiterjoin%
\definecolor{currentfill}{rgb}{1.000000,1.000000,1.000000}%
\pgfsetfillcolor{currentfill}%
\pgfsetlinewidth{1.003750pt}%
\definecolor{currentstroke}{rgb}{0.000000,0.000000,0.000000}%
\pgfsetstrokecolor{currentstroke}%
\pgfsetdash{}{0pt}%
\pgfpathmoveto{\pgfqpoint{0.701389in}{1.454132in}}%
\pgfpathlineto{\pgfqpoint{4.338485in}{1.454132in}}%
\pgfpathlineto{\pgfqpoint{4.338485in}{1.723611in}}%
\pgfpathlineto{\pgfqpoint{0.701389in}{1.723611in}}%
\pgfpathclose%
\pgfusepath{stroke,fill}%
\end{pgfscope}%
\begin{pgfscope}%
\pgfsetrectcap%
\pgfsetroundjoin%
\pgfsetlinewidth{2.007500pt}%
\definecolor{currentstroke}{rgb}{0.000000,0.000000,0.000000}%
\pgfsetstrokecolor{currentstroke}%
\pgfsetdash{}{0pt}%
\pgfpathmoveto{\pgfqpoint{0.808333in}{1.598181in}}%
\pgfpathlineto{\pgfqpoint{1.022222in}{1.598181in}}%
\pgfusepath{stroke}%
\end{pgfscope}%
\begin{pgfscope}%
\pgftext[x=1.190278in,y=1.544709in,left,base]{\sffamily\fontsize{11.000000}{13.200000}\selectfont Main}%
\end{pgfscope}%
\begin{pgfscope}%
\pgfsetbuttcap%
\pgfsetroundjoin%
\pgfsetlinewidth{2.007500pt}%
\definecolor{currentstroke}{rgb}{0.000000,0.000000,0.000000}%
\pgfsetstrokecolor{currentstroke}%
\pgfsetdash{{6.000000pt}{6.000000pt}}{0.000000pt}%
\pgfpathmoveto{\pgfqpoint{1.896606in}{1.598181in}}%
\pgfpathlineto{\pgfqpoint{2.110495in}{1.598181in}}%
\pgfusepath{stroke}%
\end{pgfscope}%
\begin{pgfscope}%
\pgftext[x=2.278551in,y=1.544709in,left,base]{\sffamily\fontsize{11.000000}{13.200000}\selectfont Intermediate}%
\end{pgfscope}%
\begin{pgfscope}%
\pgfsetbuttcap%
\pgfsetroundjoin%
\pgfsetlinewidth{2.007500pt}%
\definecolor{currentstroke}{rgb}{0.400000,0.400000,0.400000}%
\pgfsetstrokecolor{currentstroke}%
\pgfsetdash{{1.000000pt}{3.000000pt}}{0.000000pt}%
\pgfpathmoveto{\pgfqpoint{3.568240in}{1.598181in}}%
\pgfpathlineto{\pgfqpoint{3.782129in}{1.598181in}}%
\pgfusepath{stroke}%
\end{pgfscope}%
\begin{pgfscope}%
\pgftext[x=3.950184in,y=1.544709in,left,base]{\sffamily\fontsize{11.000000}{13.200000}\selectfont Title}%
\end{pgfscope}%
\end{pgfpicture}%
\makeatother%
\endgroup%

%% file: concl.tex
\section{Conclusion} 
\label{sec.conclusion}

In this work, we presented an approach to the new and challenging task of complex answer retrieval. Our approach characterizes question facets by modifying a generic neural IR architecture. We explored both approaches that focus on matching (\textit{heading independence}), and score combination (\textit{contextual vectors}).
When evaluating on the TREC CAR dataset, we achieve the top results---up to a 26\% improvement over the next best method. Furthermore, our approach significantly outperforms a leading neural IR model when evaluating with both automatic and manual judgments.
